\documentclass[journal,9pt]{IEEEtran}
\usepackage{cite}
\usepackage{amsmath,amssymb,amsfonts}
\usepackage{algorithmic}
\usepackage{algorithm}
\usepackage{graphicx}
\usepackage{textcomp}
\usepackage{xcolor}

\usepackage{booktabs}
\usepackage{siunitx}

\usepackage{comment}

\usepackage[
  colorlinks=true,
  linkcolor=blue,
  citecolor=blue,
  urlcolor=blue
]{hyperref}

\begin{document}

\title{\huge{M{\"o}bius Reparametrization of Multiport-Network Models of PIN-Diode-Programmable Metasurfaces for \\Accurate Low-Order Neumann Approximations}}

\author{\Large{Philipp del Hougne,~\IEEEmembership{Member,~IEEE}}

\thanks{
P.~del Hougne is with Univ Rennes, CNRS, IETR - UMR 6164, F-35000, Rennes, France. (e-mail: philipp.del-hougne@univ-rennes.fr)
}
\thanks{This work was supported in part by projects ANR-22-PEFT-0005 and ANR-22-CE93-0010.}

}

\maketitle

\begin{abstract}
Accurate models of programmable metasurfaces based on multiport-network theory (MNT) account for mutual coupling (MC) through a configuration-dependent matrix inversion. The latter's high computational cost during
gradient-based optimization (GBO) can be alleviated via a finite-order Neumann approximation. We show that the
accuracy of this approximation
depends strongly on the (tacitly) chosen MNT parametrization. For
1-bit-programmable meta-elements (e.g., meta-elements based on PIN diodes), we identify a closed-form M{\"o}bius reparametrization that
depends only on the meta-element's two load states and requires
neither training data nor numerical optimization.
For an experimentally estimated proxy MNT model of a fabricated 96-element
19-GHz dynamic metasurface antenna with strong MC, a second-order Neumann approximation with our reparametrization achieves forward and control-gradient accuracies of \(21.68\) and \(21.70\) dB, respectively. Relative to the full proxy MNT, it reduces the runtime of a combined forward and control-gradient evaluation by a third and the
saved-tensor memory by a factor of ten. In a prototypical GBO problem of end-to-end optimization for DMA-based
scene classification, it achieves $94.97\%$ test accuracy vs.
$95.57\%$ with the full model. 
We further note that MC-strength metrics and MC-unaware benchmarks should be parametrization-invariant, motivating, for instance,  definitions based on the best directly fitted zeroth-order model.

\end{abstract}


\section{Introduction}

Programmable metasurfaces (PMs), such as reconfigurable intelligent surfaces (RISs) and dynamic metasurface antennas (DMAs), are emerging as technological enablers of next-generation wireless systems. System models based on multiport-network theory (MNT) accurately capture the mutual coupling (MC) among the PM's tunable elements through a configuration-dependent matrix inversion~\cite{nerini2024universal,williams2022electromagnetic}. 
Consistent with its Neumann expansion, this matrix inversion can be physically interpreted as the superposition of an infinite number of multi-bounce paths between the PM's tunable elements. For the passive, lossy, and radiatively open
systems considered here, the round-trip operator
in its physical scattering parametrization has
a spectral radius below unity, ensuring convergence
of the corresponding Neumann series.

From a computational perspective, this matrix inversion can constitute a major obstacle to optimizing the PM's configuration for a desired functionality. The combination of a large number of tunable PM elements with a large number of PM configurations to evaluate during optimization can result in a prohibitive computational cost. Coordinate-descent optimization schemes can alleviate this issue through efficient low-rank inverse updates based on the Woodbury identity~\cite{prod2023efficient,prod2026updatable}. Gradient-based optimization (GBO) schemes are less amenable to such updates because every iteration generally modifies all element states and requires both a forward evaluation and an adjoint evaluation to compute the gradient. On the one hand, GBO schemes are a natural choice in the case of continuously programmable PM elements~\cite{wijekoon2024phase}. On the other hand, even when the PM elements' programmability is discrete (as in many practical prototypes), GBO can be enabled by a continuous relaxation of the discrete controls~\cite{del2020learned,MurchPixelAdjoint}; in particular, an end-to-end optimization of discretely programmable PM configurations together with downstream digital layers may rely on a GBO scheme~\cite{del2020learned,qian2022noise,delhougne2026adcaware}.
A general strategy to alleviate the cost of the matrix inversion is to replace it by a truncated Neumann expansion. Such expansions have previously been used to simplify MNT-based RIS optimization~\cite{qian2021mutual,abrardo2021mimo,mursia2023saris,ma2023ris,li2024beyond,abrardo2024design,zheng2026mutual} and RIS-based parameter estimation~\cite{zheng2024mutual,fadakar2025mutual}, but not as low-order approximations of both the forward and adjoint models needed at each GBO iteration.

To the best of our knowledge, no prior work using a truncated Neumann-series representation of an MNT model for a PM-based system has recognized that the MNT model can be reparametrized and that the finite-order truncation error depends on the chosen parametrization.
Nonetheless, the reparametrizability of the MNT model for PM-based systems has surfaced in the context of experimentally estimating the MNT parameters~\cite{delhougne2026multibit}. 
Experimentally estimating \textit{all} MNT parameters (including the characteristics of the tunable elements) is motivated by the fact that the MNT parameters for a fabricated PM-based prototype system are typically not known; reasons can include the model-reality mismatch arising from fabrication inaccuracies, the prohibitive cost of required full-wave simulations, or proprietary PM designs. Moreover, the MNT parameters usually cannot be measured directly because the PM elements are too numerous and not connectorized. Consequently, the parameters must be inferred from end-to-end measurements, for which the inverse problem is non-unique: distinct proxy MNT parameter sets can reproduce the same measurable responses. Typically, these MNT parameter ambiguities are operationally irrelevant, as they do not affect measurable quantities. However, they must be handled with care when parameter estimation is segmented across groups of PM elements~\cite{delhougne2026segmented} or across frequencies in time-Floquet systems~\cite{delhougne2026timefloquet}. Interestingly, these ambiguities have also been exploited to tighten an ambiguity-sensitive bound that depends on internal (i.e., non-end-to-end) MNT quantities~\cite{salmi2026electromagnetically}.

Here, we investigate whether reparametrizing a PM system's MNT model can improve finite-order Neumann approximations and thereby reduce the computational cost of PM optimization.
For concreteness, we consider a fabricated DMA with strong MC among its 96 PIN-diode-programmable meta-elements. Such a system with strong MC is particularly relevant for evaluating our approach, since without MC the Neumann series reduces exactly to its zeroth-order term. Moreover, strong MC can enhance a DMA's wave-domain flexibility~\cite{prodhomme2025foe,prodhomme2026benefits}, but it also complicates accurate modeling and optimization. 
\textit{First,} we derive a simple, closed-form M{\"o}bius reparametrization for PMs with 1-bit-programmable elements, and benchmark it against numerical alternatives that optimize either the reparametrization or the truncated model directly.
\textit{Second,} we formulate Neumann-truncated forward and adjoint evaluations based on this reparametrization and integrate them into end-to-end GBO. We consider the end-to-end optimization of
DMA-based scene classification~\cite{delhougne2026adcaware}, based on an experimentally calibrated MNT model for our fabricated DMA~\cite{tapie2026experimental}. Across truncation orders, we characterize the accuracy--complexity tradeoff and show that low-order GBO can achieve performance comparable to full-MNT GBO, with reduced runtime and storage.
\textit{Third,} we identify a
parametrization-invariant MC-strength metric
and a corresponding MC-unaware benchmark model.

In the broader literature on computational electromagnetics, modified Born-series formulations---including preconditioned, renormalized, learned, and change-of-variable representations---have been developed to ensure or accelerate convergence of the Neumann expansion of the Lippmann--Schwinger equation, or to improve its accuracy after only a few iterations, in strongly scattering media~\cite{kleinman1990convergent,osnabrugge2016convergent,jakobsen2016renormalized,lopezmenchon2021acceleration,stanziola2023learned,ahmadi2024physics}. In ultrasonics, such a modified Born-series solver has also been used for both forward and adjoint evaluations in full-waveform inversion~\cite{ultrasound2025}. 
These works, however, concern wave-equation
solvers for non-reconfigurable scattering
systems rather than MNT models of reconfigurable PM-based systems.
In particular, they do not consider
reconfigurable systems whose MNT models can be
reparametrized while preserving the exact
mapping from their configuration to their end-to-end transfer function. To the best of our knowledge, MNT reparametrization has not been exploited to obtain accurate fixed-low-order forward and adjoint evaluations for GBO, nor has this approach been investigated using an experimentally calibrated proxy MNT model of a fabricated PM prototype.

\section{System Model}
\label{sec:system-model}

\textit{Forward Model:} 
We consider a generic PM-based system with $N_\mathrm{T}$ input ports, $N_\mathrm{R}$ output ports, and $N_\mathrm{M}$ 1-bit-tunable lumped elements (e.g., PIN diodes). Next, we interpret each tunable lumped element as a ``virtual'' port terminated by a tunable load. Then, we partition the PM-based system into a static subsystem and a tunable subsystem, interconnected via the $N_\mathrm{M}$ ``virtual'' ports. 

The tunable subsystem comprises the loads terminating the ``virtual'' ports. The $i$th load is characterized by its reflection coefficient $r_i\in\mathbb C$. We collect the load states in
$\mathbf r=[r_1,\ldots,r_{N_\mathrm{M}}]^\mathsf{T}\in\mathbb C^{N_\mathrm{M}}$. 
For 1-bit-programmable elements, a control vector 
$\mathbf v\in\{0,1\}^{N_\mathrm{M}}$ selects between the two physical load states $\alpha,\beta\in\mathbb C$:
\begin{equation}
    \mathbf r(\mathbf v)
    =\alpha\mathbf 1_{N_\mathrm{M}}+(\beta-\alpha)\mathbf v,
    \label{eq:binary-encoding}
\end{equation}
where $\mathbf 1_{N_\mathrm M}$ denotes the
length-$N_\mathrm M$ all-ones vector. The scattering matrix of the tunable subsystem is $\mathbf\Phi(\mathbf v)\triangleq\operatorname{diag}(\mathbf r(\mathbf{v}))
\in\mathbb C^{N_\mathrm{M}\times N_\mathrm{M}}$.
The static subsystem has $N$ distinct ports, comprising the union of its input and output ports plus the $N_\mathrm{M}$ ``virtual'' ports, and is characterized by its scattering matrix $\mathbf S\in\mathbb C^{N\times N}$.

Standard MNT yields the end-to-end channel matrix $\mathbf{H}\in\mathbb{C}^{N_\mathrm{R}\times N_\mathrm{T}}$ from the $N_\mathrm{T}$ input ports to the $N_\mathrm{R}$ output ports as a function of the control vector $\mathbf{v}$:
\begin{equation}
    \mathbf H(\mathbf v)
    =
    \mathbf D+
    \mathbf A
    [\mathbf I_{N_\mathrm{M}}-\mathbf\Phi(\mathbf v)\mathbf\Gamma]^{-1}
    \mathbf\Phi(\mathbf v)\mathbf B
    \in\mathbb C^{N_\mathrm{R}\times N_\mathrm{T}},
    \label{eq:mnt-model}
\end{equation}
where $\mathbf D \triangleq\mathbf S_{\mathcal R\mathcal T} \in\mathbb C^{N_\mathrm{R}\times N_\mathrm{T}}$, $\mathbf A\triangleq\mathbf S_{\mathcal R\mathcal M} \in\mathbb C^{N_\mathrm{R}\times N_\mathrm{M}}$, $\mathbf B \triangleq\mathbf S_{\mathcal M\mathcal T}
        \in\mathbb C^{N_\mathrm{M}\times N_\mathrm{T}}$, and $\mathbf\Gamma \triangleq\mathbf S_{\mathcal M\mathcal M}
        \in\mathbb C^{N_\mathrm{M}\times N_\mathrm{M}}$. Here, $\mathcal{T}$, $\mathcal{R}$, and $\mathcal{M}$ denote the sets of port indices associated with input ports, output ports, and ``virtual'' ports, respectively; $\mathbf S_{\mathcal P\mathcal Q}$ denotes the block of $\mathbf S$
selected by index sets $\mathcal P$ and $\mathcal Q$.

\textit{Adjoint Model:}
During GBO, the binary control variables $v_i$ are relaxed to differentiable variables.
The derivative of a scalar objective
$\mathcal L(\mathbf H)\in\mathbb R$ with respect to every $v_i$ is then required. Let
$\mathbf G\in\mathbb C^{N_{\mathrm R}\times N_{\mathrm T}}$
denote the matrix derivative of $\mathcal L$ with respect to $\mathbf H$,
defined by
$\mathrm d\mathcal L
=\mathrm{Re}\,\mathrm{tr}
(\mathbf G^\dagger\mathrm d\mathbf H)$ with $(\cdot)^\dagger$ denoting the conjugate transpose. The chain rule gives
\begin{equation}
\begin{aligned}
\frac{\partial\mathcal L}{\partial v_i}
&=
\mathrm{Re}\,\mathrm{tr}\left(
\mathbf G^\dagger
\frac{\partial\mathbf H}{\partial v_i}
\right)\\
&=
\mathrm{Re}\,\mathrm{tr}\left[
\mathbf G^\dagger\mathbf A\mathbf M^{-1}
\frac{\partial\boldsymbol{\Phi}}{\partial v_i}
\left(
\mathbf B+
\boldsymbol{\Gamma}\mathbf M^{-1}
\boldsymbol{\Phi}\mathbf B
\right)
\right]\\
&=
\mathrm{Re}\,\mathrm{tr}\left[
(\beta-\alpha)
\mathbf G^\dagger\mathbf A\mathbf M^{-1}
\mathbf e_i\mathbf e_i^{\mathsf T}
\left(
\mathbf B+
\boldsymbol{\Gamma}\mathbf M^{-1}
\boldsymbol{\Phi}\mathbf B
\right)
\right],
\end{aligned}
\label{eq:direct-control-gradient}
\end{equation}
where $\mathbf M\triangleq
\mathbf I_{N_{\mathrm M}}-\boldsymbol{\Phi}\boldsymbol{\Gamma}$,
and $\mathbf e_i$ is the $i$th canonical basis vector. The second equality
follows by differentiating \eqref{eq:mnt-model}, whereas the third uses
$\partial\boldsymbol{\Phi}/\partial v_i
=(\beta-\alpha)\mathbf e_i\mathbf e_i^{\mathsf T}$. 
At a given GBO iteration, the factor $ \mathbf{\Lambda}^\dagger = \mathbf G^\dagger\mathbf A\mathbf M^{-1}$ is common to the derivative of $\mathcal{L}$ with respect to each \(v_i\). The adjoint-variable method computes $\mathbf{\Lambda}^\dagger$ once,
rather than separately for every $v_i$, and reuses it to obtain all
$N_{\mathrm M}$ components of the control gradient
\cite{Nikolova2004,veronis2004method}.

\textit{Neumann Approximation:}
\eqref{eq:mnt-model} and~\eqref{eq:direct-control-gradient} both use $\mathbf M^{-1}$, and evaluating~\eqref{eq:direct-control-gradient} via the adjoint variable
$\boldsymbol\Lambda=(\mathbf M^{-1})^\dagger\mathbf A^\dagger\mathbf G$
also uses $\left[\mathbf M(\mathbf v)^{-1}\right]^\dagger$. 
The Neumann expansions of $\mathbf M(\mathbf v)^{-1}$ and $\left[\mathbf M(\mathbf v)^{-1}\right]^\dagger$ are:
\begin{subequations}
\label{eq:forward-adjoint-neumann}
\begin{align}
\mathbf M(\mathbf v)^{-1}
&=
\sum_{k=0}^{\infty}
\bigl(
\boldsymbol{\Phi}(\mathbf v)\boldsymbol{\Gamma}
\bigr)^k,
\label{eq:forward-neumann}\\
\left[\mathbf M(\mathbf v)^{-1}\right]^\dagger
&=
\sum_{k=0}^{\infty}
\bigl(
\boldsymbol{\Gamma}^\dagger
\boldsymbol{\Phi}(\mathbf v)^\dagger
\bigr)^k.
\label{eq:adjoint-neumann}
\end{align}
\end{subequations}
For the passive, lossy, and radiatively open systems considered here,
the round-trip operator $\boldsymbol{\Phi}\boldsymbol{\Gamma}$
formed from the ``ground-truth'' MNT parameters has a spectral radius below unity. This statement refers to the set of MNT parameters that could, conceptually, be measured if the ``virtual'' ports were accessible. The adjoint $\boldsymbol{\Gamma}^{\dagger}\boldsymbol{\Phi}^{\dagger}
=(\boldsymbol{\Phi}\boldsymbol{\Gamma})^{\dagger}$
has the same spectral radius, so both series
in~\eqref{eq:forward-adjoint-neumann} converge. In contrast, an end-to-end-fitted 
proxy MNT parametrization need not preserve this property.

Truncating the series in~\eqref{eq:forward-adjoint-neumann} at order $K$ yields the approximations
\begin{subequations}
\label{eq_trunc}
\label{eq:truncations}
\begin{align}
\mathbf H_K(\mathbf v)
&=
\mathbf D+
\mathbf A
\left[
\sum_{k=0}^{K}
\bigl(
\boldsymbol{\Phi}(\mathbf v)\boldsymbol{\Gamma}
\bigr)^k
\right]
\boldsymbol{\Phi}(\mathbf v)\mathbf B,
\label{eq:neumann-truncation}\\
\boldsymbol{\Lambda}_K(\mathbf v)
&=
\left[
\sum_{k=0}^{K}
\bigl(
\boldsymbol{\Gamma}^\dagger
\boldsymbol{\Phi}(\mathbf v)^\dagger
\bigr)^k
\right]
\mathbf A^\dagger\mathbf G_K(\mathbf v),
\label{eq:adjoint-neumann-truncation}
\end{align}
\end{subequations}
where $\mathbf G_K(\mathbf{v})$ denotes the derivative of
the loss with respect to $\mathbf H$, evaluated
at the truncated forward response
$\mathbf H_K(\mathbf v)$.
The $k$th forward term
in~\eqref{eq:neumann-truncation} represents
paths involving $k$ rescattering events at ``virtual'' ports, and hence $k+1$ interactions with the tunable loads. Both sums are evaluated recursively without matrix inversions.
We distinguish the \textit{unrolled} gradient, obtained by differentiating the finite-order forward model ${\mathbf H}_K$, from our \textit{implicit} approximate adjoint: the latter evaluates the exact gradient identity \eqref{eq:direct-control-gradient} using the order-$K$ forward and adjoint Neumann approximations in~\eqref{eq_trunc}. 
Thus, the unrolled and implicit gradients need not coincide.

\textit{Reparametrization:} 
As described in [Appendix~A,~\cite{del2026electromagnetic}], there are three reparametrizations of the MNT model in~\eqref{eq:mnt-model} that leave the $\mathbf{v}\rightarrow\mathbf{H}(\mathbf{v})$ mapping unaffected. Of these three, only the M{\"o}bius transformation alters the
individual terms in~\eqref{eq_trunc}, thereby changing the
finite-order forward and adjoint approximations
$\mathbf H_K(\mathbf v)$ and $\boldsymbol{\Lambda}_K(\mathbf v)$,
and the associated truncation errors in the forward response
and control gradient. Denoting transformed parameters with a tilde, the M{\"o}bius transformation of the MNT model yields
\begin{equation}
\begin{aligned}
    \widetilde\alpha&=\mathcal M_m(\alpha), &
    \widetilde\beta&=\mathcal M_m(\beta), &
    \widetilde{\mathbf D}&=\mathbf D+m\mathbf A\mathbf F\mathbf B,\\
    \widetilde{\mathbf A}&=\kappa\mathbf A\mathbf F, &
    \widetilde{\mathbf B}&=\kappa\mathbf F\mathbf B, &
    \widetilde{\mathbf\Gamma}&=(\mathbf\Gamma-m^*\mathbf I_{N_\mathrm M})\mathbf F ,
\end{aligned}
\label{eq:mobius-reparametrization}
\end{equation}
where $\mathbf F\triangleq
(\mathbf I_{N_\mathrm M}-m\mathbf\Gamma)^{-1}$,
$\kappa\triangleq\sqrt{1-|m|^2}$, and
$\mathcal M_m(\rho)\triangleq
(\rho-m)/(1-m^*\rho)$.
We denote by $\mathcal{A}$ the admissible 
set of M{\"o}bius parameters $m\in\mathbb C$
satisfying $|m|<1$,
$\mathbf I_{N_\mathrm M}-m\mathbf\Gamma$
nonsingular, and $1-m^*\rho\neq0$ for
$\rho\in\{\alpha,\beta\}$.
For real $m$, this transformation corresponds to changing the common real-valued reference impedance at the ``virtual'' ports; complex values of $m$ provide a more general M{\"o}bius reparametrization.
For relaxed control variables, we use
$\widetilde r_i(v_i)
\triangleq\mathcal M_m(r_i(v_i))$ and
$\widetilde{\mathbf\Phi}(\mathbf v)
\triangleq
\operatorname{diag}
(\widetilde{\mathbf r}(\mathbf v))$.
Accordingly,
$\partial\widetilde{\boldsymbol{\Phi}}
/\partial v_i
=
\kappa^2(\beta-\alpha)
[1-m^*r_i(v_i)]^{-2}
\mathbf e_i\mathbf e_i^{\mathsf T}$.

\section{M{\"o}bius Reparametrization}
\label{sec:reparametrization}

Although the exact forward response in \eqref{eq:mnt-model} and the exact
derivatives with respect to the control variables in
\eqref{eq:direct-control-gradient} are invariant under the M{\"o}bius
reparametrization, the corresponding order-$K$ approximations based on
\eqref{eq:truncations} are not. We therefore seek a M{\"o}bius
reparametrization that reduces their finite-order truncation errors.

\textit{Closed Form (CF):}
For a binary configuration, $\widetilde{\mathbf\Phi}(\mathbf v)$ is
diagonal and its $i$th diagonal entry is either
$\widetilde\alpha=\mathcal M_m(\alpha)$ or $\widetilde\beta=\mathcal M_m(\beta)$. Hence, for an admissible M{\"o}bius
reparametrization, the spectral norm of  $\widetilde{\mathbf\Phi}(\mathbf v)$ satisfies
\begin{equation}
    \|\widetilde{\mathbf\Phi}(\mathbf v)\|_2
    \leq
    q(m)
    \triangleq
    \max_{\rho\in\{\alpha,\beta\}}
    |\mathcal M_m(\rho)|.
    \label{eq7}
\end{equation}
For binary configurations, substituting the
transformed parameters from
\eqref{eq:mobius-reparametrization} into
\eqref{eq:truncations} yields the following
bounds on the $k$th Neumann terms:
\begin{subequations}
\label{eq:term-bounds}
\begin{align}
\left\|
\widetilde{\mathbf A}
(\widetilde{\mathbf\Phi}\widetilde{\mathbf\Gamma})^k
\widetilde{\mathbf\Phi}\widetilde{\mathbf B}
\right\|_{\mathrm F}
&\leq
\|\widetilde{\mathbf A}\|_2
\|\widetilde{\mathbf B}\|_{\mathrm F}
q(m)^{k+1}
\|\widetilde{\mathbf\Gamma}\|_2^k,
\label{eq:forward-term-bound}\\
\left\|
(\widetilde{\mathbf\Gamma}^{\dagger}
\widetilde{\mathbf\Phi}^{\dagger})^k
\widetilde{\mathbf A}^{\dagger}
\widetilde {\mathbf G}_K(\mathbf v)
\right\|_{\mathrm F}
&\leq
q(m)^k
\|\widetilde{\mathbf\Gamma}\|_2^k
\|\widetilde{\mathbf A}^{\dagger}
\widetilde{\mathbf G}_K(\mathbf v)\|_{\mathrm F},
\label{eq:adjoint-term-bound}
\end{align}
\end{subequations}
where $\widetilde{\mathbf G}_K(\mathbf v)$
denotes the derivative of the loss with respect to $\mathbf{H}$, evaluated at $\widetilde{\mathbf H}_K(\mathbf v)$, and we use
$\|\mathbf X\mathbf Y\|_{\mathrm F}
\leq\|\mathbf X\|_2\|\mathbf Y\|_{\mathrm F}$ and $\|\widetilde{\mathbf\Phi}^{\dagger}\|_2
=\|\widetilde{\mathbf\Phi}\|_2$ in combination with~\eqref{eq7}.
Although the other factors in \eqref{eq:term-bounds} also depend on $m$,
$q(m)$ is the only factor determined solely by the load states and appears once per load interaction.
Assuming $|\alpha|,|\beta|<1$, as expected for
lossy passive load states, this motivates choosing the CF reparametrization to minimize $q(m)$
uniformly over all binary configurations:
\begin{equation}
    m_{\mathrm{CF}} (\alpha,\beta)
    \triangleq
    \arg\min_{|m|<1}q(m) = \frac{\alpha+w(\alpha,\beta)}{1+\alpha^*w(\alpha,\beta)},
    \label{eq:closed-form-objective}
\end{equation}
where $w=d/(1+\sqrt{1-|d|^2})$ and $d=(\beta-\alpha)/(1-\alpha^*\beta)$. The closed-form solution in~\eqref{eq:closed-form-objective} is simply the hyperbolic midpoint of $\alpha$ and $\beta$, implying $\mathcal M_{m_\mathrm{CF}}(\alpha) = - \mathcal M_{m_\mathrm{CF}}(\beta)$. 
Provided that $m_\mathrm{CF}\in\mathcal{A}$,
substituting it into
\eqref{eq:mobius-reparametrization} and then
into \eqref{eq_trunc} yields our proposed
approximations.

\textit{Numerical Benchmarks:}
For a training set $\mathcal V$ of control
vectors, we define the forward- and adjoint-optimal
M{\"o}bius parameters for a given order $K$ by
\begin{subequations}
\label{eq:numerical-benchmarks}
\begin{equation}
m_{K,\mathrm{fwd}}
\in
\underset{m\in\mathcal A}{\operatorname{arg\,min}}\;
\frac{
\sum_{\mathbf v\in\mathcal V}
\left\|
\widetilde{\mathbf H}_K(\mathbf v;m)-\mathbf H(\mathbf v)
\right\|_{\mathrm F}^2
}{
\sum_{\mathbf v\in\mathcal V}
\left\|\mathbf H(\mathbf v)\right\|_{\mathrm F}^2
},
\label{eq:response-oracle}
\end{equation}
\begin{equation}
m_{K,\mathrm{adj}}
\in
\underset{m\in\mathcal A}{\operatorname{arg\,min}}\;
\frac{
\sum_{\mathbf v_j\in\mathcal V}
\left\|
\mathbf g^{\mathrm{imp}}_{K,j}(\mathbf v_j;m)
-\mathbf g_j(\mathbf v_j)
\right\|_2^2
}{
\sum_{\mathbf v_j\in\mathcal V}
\left\|\mathbf g_j(\mathbf v_j)\right\|_2^2
}.
\label{eq:adjoint-oracle}
\end{equation}
\end{subequations}
In~\eqref{eq:adjoint-oracle},
$\mathbf g_j(\mathbf v)
\triangleq\nabla_{\mathbf v}
\mathcal L_j(\mathbf H(\mathbf v))$
is the full-MNT control gradient, while
$\mathbf g^{\mathrm{imp}}_{K,j}(\mathbf v;m)$ is its order-$K$
estimate computed with the truncated forward model and the truncated implicit
adjoint.
A control gradient is a vector--Jacobian product (VJP): it
propagates a loss sensitivity at the response back to the DMA controls.
To assess the control-gradient accuracy independently of the downstream
digital classifier, we draw for each
$\mathbf v_j\in\mathcal V$ a random matrix
$\mathbf G_j\in\mathbb C^{N_{\mathrm R}\times N_{\mathrm T}}$
with $\|\mathbf G_j\|_{\mathrm F}=1$ 
and define the auxiliary scalar loss
$\mathcal L_j(\mathbf Y)
=\operatorname{Re}\operatorname{tr}(\mathbf G_j^\dagger\mathbf Y)$
for $\mathbf Y\in\mathbb C^{N_{\mathrm R}\times N_{\mathrm T}}$.
We evaluate $\mathcal L_j(\mathbf Y)$ at $\mathbf Y=\mathbf H(\mathbf v)$ or
$\mathbf Y=\widetilde{\mathbf H}_K(\mathbf v;m)$.
Because $\mathcal L_j(\mathbf Y)$ is linear in $\mathbf{Y}$, its matrix derivative is the same
$\mathbf G_j$ at the full and truncated responses. We keep
$\mathbf G_j$ fixed as $m$ varies, so that the control-gradient
difference in the numerator of~\eqref{eq:adjoint-oracle} probes the MNT gradient approximation rather than a
response-dependent change in the loss sensitivity.
The M{\"o}bius parameters $m_{K,\mathrm{fwd}}$ and
$m_{K,\mathrm{adj}}$ serve as separate numerical benchmarks for
forward-response and control-gradient errors and are not combined
during GBO. We estimate them by a coarse search followed by
multistart Nelder--Mead optimization.

As a less restricted $K=0$ benchmark, we directly fit a
zeroth-order truncation of the MNT model:
\begin{equation}
\widehat{\mathbf H}(\mathbf v;\boldsymbol\vartheta_0)
=
\mathbf C+
\sum_{n=1}^{N_{\mathrm M}}
v_n\mathbf u_n\mathbf z_n^{\mathsf T},
\label{eq:direct-k0}
\end{equation}
where $\boldsymbol\vartheta_0
\triangleq
\{\mathbf C,\mathbf u_n,\mathbf z_n\}_{n=1}^{N_{\mathrm M}}$ are the model parameters and each single-element contribution has rank at most one. We define the optimal parameter set as 
\begin{equation}
\widehat{\boldsymbol\vartheta}_{0,\mathrm{dir}}
\in
\underset{\boldsymbol\vartheta_0}{\operatorname{arg\,min}}\;
\frac{
\sum_{\mathbf v\in\mathcal V}
\left\|
\widehat{\mathbf H}(\mathbf v;\boldsymbol\vartheta_0)
-\mathbf H(\mathbf v)
\right\|_{\mathrm F}^2}
{
\sum_{\mathbf v\in\mathcal V}
\left\|\mathbf H(\mathbf v)\right\|_{\mathrm F}^2}.
\label{eq:direct-k0-fit}
\end{equation}
We initialize our estimate of
$\widehat{\boldsymbol\vartheta}_{0,\mathrm{dir}}$ by complex least
squares followed by rank-one SVD projection\footnote{Because under MC each single-flip response  generally depends on the background configuration, least squares estimates one unrestricted,
configuration-independent affine coefficient per element from all
training samples; the SVD then provides its closest rank-one
approximation.}; we refine our estimate with full-batch gradient-based optimization of
the normalized response mean-squared error.
Unlike the M{\"o}bius-constrained MNT reparametrizations, we fit all parameters of the affine model in~\eqref{eq:direct-k0} freely, making it a stringent $K=0$ benchmark.

\section{Results}

We now evaluate the M{\"o}bius reparametrizations from Sec.~\ref{sec:reparametrization} in the context of a concrete sensing task, based on experimentally estimated proxy MNT parameters.

\begin{figure}[t]
\centering
\includegraphics[width=\columnwidth]{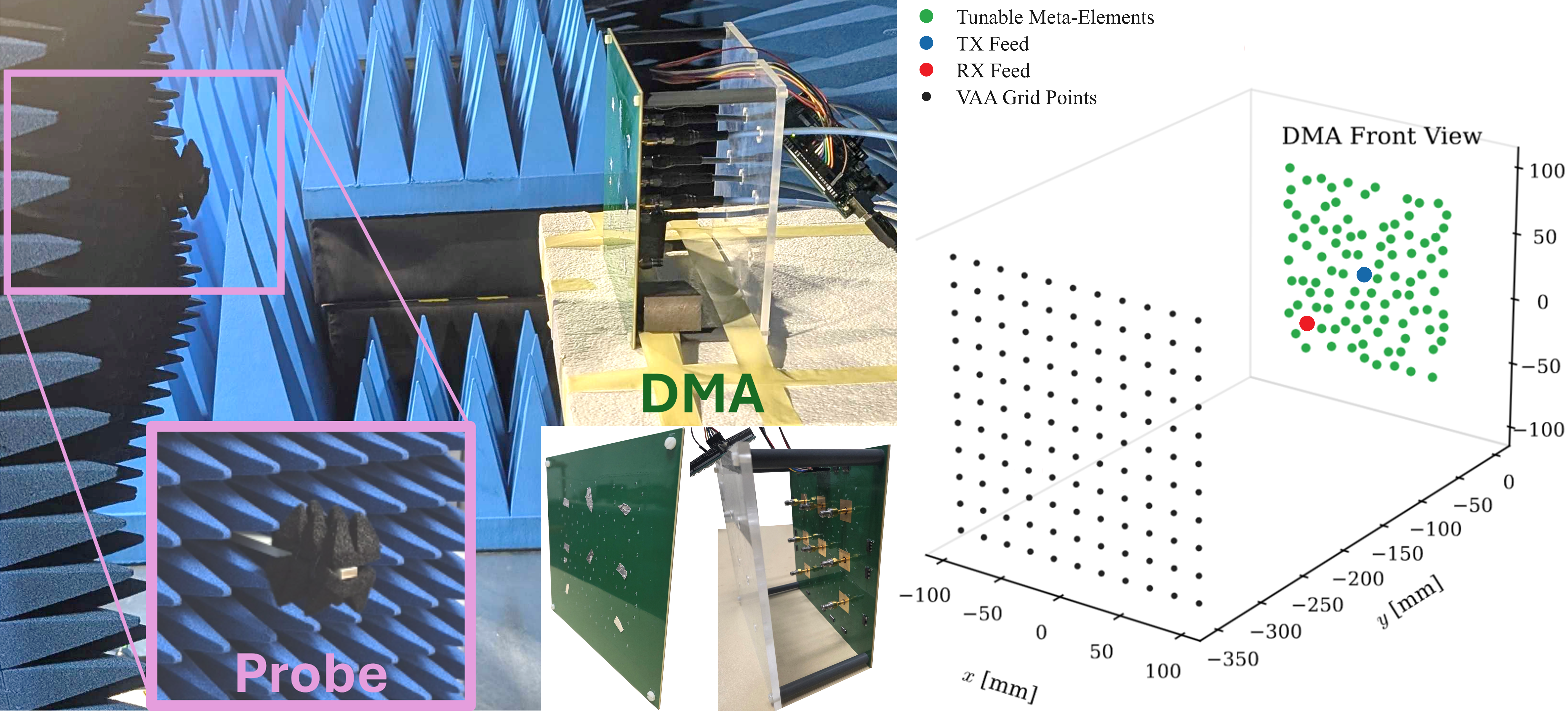}
\caption{Experimental setup; see~\cite{tapie2026experimental} for details.}
\label{Fig1}
\end{figure}

\textit{Setup and Initial Proxy MNT Model:}
We use the experimentally calibrated proxy MNT model of the multi-feed 19-GHz DMA in~\cite{tapie2026experimental}. Its 96 PIN-diode-programmable meta-elements and eight feeds are strongly coupled to each other via a quasi-2D chaotic cavity. The design follows that of a similar
single-feed DMA in~\cite{sleasman2020implementation}. We sample the field radiated by the DMA using a virtual antenna array (VAA) with $N_\mathrm{G}=121$ regularly spaced grid points ($2$-cm step,
in a plane parallel to and $35$\,cm in front of the DMA), as shown in
Fig.~\ref{Fig1}. 
The DMA's feed-to-VAA transmission matrix $\mathbf T(\mathbf v)$
follows \eqref{eq:mnt-model}, while its feed-to-feed scattering matrix
$\mathbf R(\mathbf v)$ has the same MNT form
\cite{tapie2026experimental,delhougne2026adcaware}. The two models share $\mathbf B$, $\mathbf\Gamma$, $\alpha$, and $\beta$, whereas $\mathbf D$ and $\mathbf A$ differ. 
We estimated the initial proxy MNT model (identified by $^{(0)}$) imposing $\widetilde\alpha^{(0)} = 0$ and reciprocity, but without imposing passivity, as described in~\cite{tapie2026experimental}. While this parametrization accurately predicts $\mathbf T(\mathbf v)$ and $\mathbf R(\mathbf v)$, its Neumann series need not converge; for instance, for the all-$\widetilde{\beta}^{(0)}$
configuration, the spectral radius of $\widetilde{\boldsymbol\Phi}^{(0)}\widetilde{\mathbf\Gamma}^{(0)}$ is $2.03>1$.

\textit{Sensing Task:}
The monostatic scene-classification task follows
\cite{delhougne2026adcaware}: one DMA feed transmits, another DMA feed receives, and the scene reflectivity vector $\boldsymbol\rho\in[0,1]^{N_\mathrm{G}}$ encodes an MNIST image. The DMA's six unused feeds are open-circuited.\footnote{One feed is physically open-circuited; the other five are physics-consistently eliminated via MNT reduction with open-circuit terminations~\cite{delhougne2026adcaware}.}
We denote by  $\mathbf t_{\rm tx}(\mathbf v_j)\in\mathbb{C}^{N_\mathrm{G}}$, $\mathbf t_{\rm rx}(\mathbf v_j)\in\mathbb{C}^{N_\mathrm{G}}$, and $s(\mathbf v_j)\in\mathbb{C}$ the resulting TX-VAA, RX-VAA, and TX-RX transmission, respectively. 
Under the common approximation of a single scattering event within the scene, the complex-valued scalar measurement is $z_j^{\rm pre}(\boldsymbol\rho)
  =\boldsymbol\rho^{\mathsf T}\mathbf p(\mathbf v_j)+s(\mathbf v_j) \in\mathbb{C}$, where $\mathbf p(\mathbf v_j)\triangleq
\mathbf t_{\rm rx}(\mathbf v_j)\odot\mathbf t_{\rm tx}(\mathbf v_j)
\in\mathbb C^{N_\mathrm{G}}$ with $\odot$ denoting elementwise multiplication~\cite{delhougne2026adcaware}. Assuming model-based ideal self-interference cancellation, the complex-valued scalar measurement becomes $z_j(\boldsymbol\rho)  =z_j^{\rm pre}(\boldsymbol\rho)-s(\mathbf v_j) = \boldsymbol\rho^{\mathsf T}\mathbf p(\mathbf v_j)$~\cite{delhougne2026adcaware}. For simplicity, we assume negligible measurement noise and an ideal analog-to-digital conversion. We optimize a sequence of $M$ DMA configurations jointly with a downstream digital classifier. Further details can be found in~\cite{delhougne2026adcaware}. We deviate from~\cite{delhougne2026adcaware} only regarding the use of truncated forward and adjoint models for training and validation. We always evaluate final test accuracies based on the full proxy MNT model.

\textit{M{\"o}bius Reparametrizations:}
Our initial proxy load states satisfy $|\widetilde{\alpha}^{(0)}|,|\widetilde{\beta}^{(0)}|<1$.
We can uniquely define $m_\mathrm{CF}$ based on $\widetilde{\alpha}^{(0)}$ and $\widetilde{\beta}^{(0)}$. 
We can separately apply the numerical objectives in~\eqref{eq:numerical-benchmarks} to $\mathbf T$ or $\mathbf R$ by setting
$\mathbf H=\mathbf T$ or $\mathbf H=\mathbf R$; for the relevant sensing variables collected in $\mathbf y(\mathbf v)\triangleq
[\mathbf p(\mathbf v)^{\mathsf T},s(\mathbf v)]^{\mathsf T}$, we
can instead minimize the analogous forward-response and control-gradient errors of $\mathbf y$, yielding $m_{K,\mathrm{fwd}}^{(\mathrm y)}$ and
$m_{K,\mathrm{adj}}^{(\mathrm y)}$.
For the direct $K=0$ benchmark, we jointly fit~\eqref{eq:direct-k0} to measured $\mathbf T$ and $\mathbf R$ with equal variance-normalized mean-squared-error weights. The offsets and per-element left factors are separate, while the right factors are shared; $\mathbf y$ is then constructed from the fit.
In contrast to the initial proxy MNT parametrization, the Neumann series converges for the all-$\widetilde{\beta}$-configuration after reparametrization with $m_\mathrm{CF}$, $m_{K,\mathrm{fwd}}^{(\mathrm y)}$,
or $m_{K,\mathrm{adj}}^{(\mathrm y)}$, yielding a spectral radius of $\widetilde{\boldsymbol\Phi} \widetilde{\mathbf\Gamma}$ between $0.82$ and $0.92$.

\textit{Truncated-Model Accuracy:}
We quantify the forward-model accuracy for
$\mathbf X\in\{\mathbf T,\mathbf R\}$ using\footnote{
Unlike~\cite{tapie2026experimental}, we take the ratio of
entry-averaged signal and residual standard deviations rather than
averaging entrywise ratios. This limits the influence of entries with
nearly constant residuals.}
\begin{equation}
\zeta_{\mathrm X}
=20\log_{10}\left(
\frac{\bar\sigma(\mathbf X_{\rm ref})}
     {\bar\sigma(\mathbf X_{\rm ref}-\mathbf X_{\rm pred})}
\right),
\label{zeta_definition}
\end{equation}
where $\bar\sigma(\mathbf X)\triangleq
\langle\operatorname{SD}_\ell(X_{ij}^{(\ell)})\rangle_{i,j}$
and $\operatorname{SD}_\ell$ is taken over unseen held-out control vectors.
$\zeta_{\mathrm X}$ compares the configuration-dependent variation of the reference with that of the prediction residual, akin to a signal-to-noise ratio.

\begin{figure}[t]
\centering
\includegraphics[width=\columnwidth]{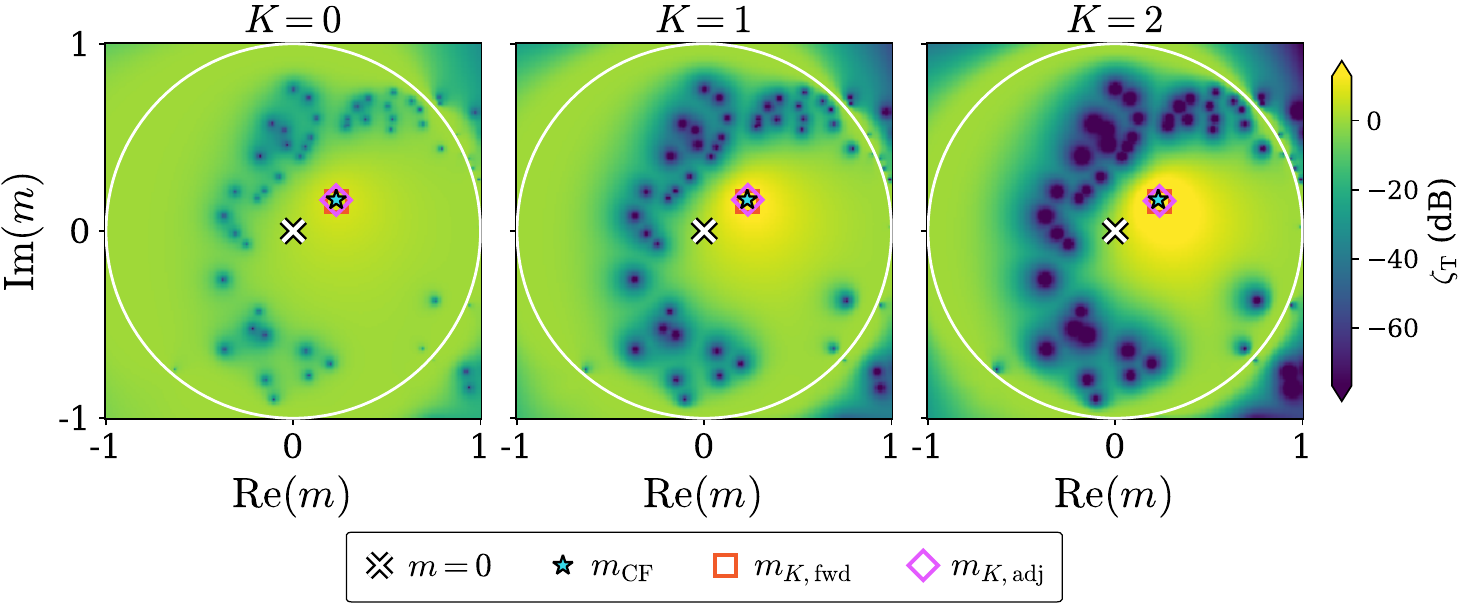}
\caption{Accuracy of the order-$K$ truncated $\mathbf T$ model (relative to the full proxy MNT) vs. $m$. The white circles mark $|m|=1$; our optimizations use $m\in\mathcal A$.}\label{FigM}
\end{figure}

We first evaluate the accuracy of the order-$K$ truncated
$\mathbf T(\mathbf v)$ model from~\eqref{eq:neumann-truncation}. While the full proxy MNT model is by definition insensitive to $m$, our results for $K\in\{0,1,2\}$ in Fig.~\ref{FigM} display a strong dependence on $m$ for the accuracy of the order-$K$ truncated model. $m_\mathrm{CF}$ visually coincides with $m_{K,\mathrm{fwd}}$ and $m_{K,\mathrm{adj}}$, corroborating the practical utility of $m_\mathrm{CF}$. 
Both observations also hold qualitatively for the forward-model
error of $\mathbf y$ and for the
control-gradient approximation errors associated with both
$\mathbf T$ and $\mathbf y$.\footnote{Our VJP tests for $\mathbf T$ use the random-matrix sensitivities $\mathbf G_j$ defined above. For $\mathbf y$, we
analogously draw a unit-norm random complex vector $\mathbf{a}_j\in\mathbb C^{N_{\mathrm G}+1}$ for each $\mathbf{v}_j$, held fixed across our exact--approximate comparisons. The chain rule propagates $\mathbf a_j$ through $\mathbf p$, $s$,
and the open-circuit reduction to the relevant entries of $\mathbf T$
and $\mathbf R$.}

The near-optimality of $m_{\rm CF}$ is further corroborated in Table~\ref{tab:raw_zeta} which shows that $m_{\rm CF}$ achieves nearly the same forward-model accuracies as $m_{K,\mathrm{fwd}}^\mathrm{(y)}$ at each $K$. Moreover, the accuracy with  $m_{\rm CF}$ at $K=0$ is comparable to the independently fitted rank-one models of~\eqref{eq:direct-k0}. Meanwhile, without reparametrization ($m=0$), the convergence of the Neumann series is uncertain. Correspondingly, its $K\leq2$ truncations have essentially
no predictive value.

\begin{table}[t]
\caption{Forward-model accuracies $\zeta_\mathrm{R}$ and $\zeta_\mathrm{T}$ relative to held-out measurements, reported as $\zeta_\mathrm{R} \ | \ \zeta_\mathrm{T}$ in dB.}
\label{tab:raw_zeta}
\centering
\scriptsize
\setlength{\tabcolsep}{1pt}
\begin{tabular}{@{}l%
S[table-format=-2.2]@{\hspace{1pt}\textbar\hspace{1pt}}S[table-format=2.2]@{\hspace{3pt}}%
S[table-format=-2.2]@{\hspace{1pt}\textbar\hspace{1pt}}S[table-format=2.2]@{\hspace{3pt}}%
S[table-format=-2.2]@{\hspace{1pt}\textbar\hspace{1pt}}S[table-format=2.2]@{\hspace{3pt}}%
S[table-format=-2.2]@{\hspace{1pt}\textbar\hspace{1pt}}S[table-format=2.2]@{}}
\toprule
Model
& \multicolumn{2}{c}{$K=0$}
& \multicolumn{2}{c}{$K=1$}
& \multicolumn{2}{c}{$K=2$}
& \multicolumn{2}{c}{Full MNT} \\
\midrule
$m=0$
& -0.50 & 2.17 & -1.58 & 3.09 & -3.13 & 2.93 & 37.95 & 37.35 \\
$m_{\rm CF}$
&  7.08 & 7.11 & 13.53 & 13.61 & 19.82 & 20.09 & 37.95 & 37.35 \\
$m_{K,\mathrm{fwd}}^{(\mathrm{y})}$
&  6.94 & 7.04 & 13.51 & 13.58 & 19.65 & 19.91 & 37.95 & 37.35 \\
\midrule
Joint $\mathbf{T}$ and $\mathbf{R}$ fit
& 6.86 & 5.89
& \multicolumn{2}{c}{N/A}
& \multicolumn{2}{c}{N/A}
& \multicolumn{2}{c}{N/A} \\
$\mathbf T$-only fit
& \multicolumn{1}{r@{\hspace{1pt}\textbar\hspace{1pt}}}{N/A} & 5.81
& \multicolumn{2}{c}{N/A}
& \multicolumn{2}{c}{N/A}
& \multicolumn{2}{c}{N/A} \\
$\mathbf R$-only fit
& 6.98 & {N/A}
& \multicolumn{2}{c}{N/A}
& \multicolumn{2}{c}{N/A}
& \multicolumn{2}{c}{N/A} \\
\bottomrule
\end{tabular}
\end{table}

Next, we compare in Table~\ref{tab:physical_cost} the accuracy and computational cost of evaluating $\mathbf y$ and its control VJP.\footnote{Measured on two Intel
Xeon Gold 6248R CPUs (48 threads, PyTorch 2.8.0), with all $M=128$
configurations in one batch. Times are medians of 15 runs after
three warm-ups, excluding the reparametrization. ``Act.'' measures autograd-saved
tensor memory, excluding fixed or cached MNT tensors.}
We obtain $\zeta_{\mathbf y}$ by applying~\eqref{zeta_definition}
to the entries of $\mathbf y$. For $\zeta_{\mathrm{VJP}}$, we apply
the same metric to the components of the control gradients of
$\operatorname{Re}[\mathbf a_j^\dagger\mathbf y(\mathbf v)]$ at
$\mathbf v_j$, using the same $\mathbf a_j$ for the full and
approximate models at each configuration.
With $m_\mathrm{CF}$, the forward and VJP accuracies are closely
matched at each order: at $K=2$, they reach 21.68 and 21.70\,dB,
respectively, close to the accuracies obtained with $m_{K,\mathrm{fwd}}^{\mathrm{(y)}}$ and $m_{K,\mathrm{adj}}^{\mathrm{(y)}}$. The forward-plus-VJP evaluation time falls from 14.32 to 9.09\,ms
and autograd-saved tensor memory from 19.72 to 1.95\,MiB relative to full MNT.
At $K=0$, the direct rank-one fit has comparable forward accuracy
to $m_\mathrm{CF}$ (7.26 vs.\ 7.52\,dB) but lower VJP accuracy
(4.96 vs.\ 7.52\,dB). Unrolling yields lower VJP accuracy than
the implicit adjoint at every $K$.

\newcommand{\unr}[1]{\textcolor{black!60}{[#1]}}

\newcommand{\padone}[1]{\phantom{0}#1}
\newcommand{\blankunr}[1]{#1 \phantom{\unr{00.00}}}

\begin{table}[t]
\caption{Accuracies of $\mathbf y$ and its control VJP relative to the
full proxy MNT over 1,280 configurations, with computational cost at
$M=128$.  Time includes forward evaluation and VJP;
Act. is saved-tensor memory. imp: implicit; unr: unrolled.}
\label{tab:physical_cost}
\centering\scriptsize
\setlength{\tabcolsep}{2.3pt}
\renewcommand{\arraystretch}{1.05}
\begin{tabular}{@{}lcrrrr@{}}
\toprule
Model & $K$ & $\zeta_{\mathbf y}$ (dB) &
$\zeta_{\mathrm{VJP}}$ (dB) & Time (ms) & Act. (MiB) \\
\midrule
Full MNT & N/A & $\infty$ &
\blankunr{$\infty$} & \blankunr{14.32} & \blankunr{19.72} \\
\midrule
$m_\mathrm{CF}$ (imp. \unr{unr.}) & 0 &
\padone{7.52} & \padone{7.52} \unr{\padone{5.36}} &
\padone{6.34} \unr{\padone{5.81}} &
\padone{1.95} \unr{\padone{1.20}} \\
$m_\mathrm{CF}$ (imp. \unr{unr.}) & 1 &
14.87 & 14.71 \unr{10.98} &
\padone{7.82} \unr{\padone{7.48}} &
\padone{1.95} \unr{\padone{1.95}} \\
$m_\mathrm{CF}$ (imp. \unr{unr.}) & 2 &
21.68 & 21.70 \unr{16.68} &
\padone{9.09} \unr{\padone{8.99}} &
\padone{1.95} \unr{\padone{2.61}} \\
\midrule
$m_{K,\mathrm{fwd}}^{(\mathrm y)}$ (imp. \unr{unr.}) & 0 &
\padone{7.34} & \padone{7.39} \unr{\padone{5.23}} &
\padone{6.51} \unr{\padone{6.04}} &
\padone{1.95} \unr{\padone{1.20}} \\
$m_{K,\mathrm{fwd}}^{(\mathrm y)}$ (imp. \unr{unr.}) & 1 &
14.75 & 14.63 \unr{10.91} &
\padone{8.04} \unr{\padone{7.79}} &
\padone{1.95} \unr{\padone{1.95}} \\
$m_{K,\mathrm{fwd}}^{(\mathrm y)}$ (imp. \unr{unr.}) & 2 &
21.36 & 21.42 \unr{16.37} &
\padone{9.19} \unr{\padone{9.55}} &
\padone{1.95} \unr{\padone{2.61}} \\
\midrule
$m_{K,\mathrm{adj}}^{(\mathrm y)}$ (imp. \unr{unr.}) & 0 &
\padone{7.55} & \padone{7.53} \unr{\padone{5.37}} &
\padone{6.50} \unr{\padone{6.11}} &
\padone{1.95} \unr{\padone{1.20}} \\
$m_{K,\mathrm{adj}}^{(\mathrm y)}$ (imp. \unr{unr.}) & 1 &
14.89 & 14.72 \unr{10.98} &
\padone{7.77} \unr{\padone{7.80}} &
\padone{1.95} \unr{\padone{1.95}} \\
$m_{K,\mathrm{adj}}^{(\mathrm y)}$ (imp. \unr{unr.}) & 2 &
21.68 & 21.68 \unr{16.68} &
\padone{8.85} \unr{\padone{9.50}} &
\padone{1.95} \unr{\padone{2.61}} \\
\midrule
Joint $\mathbf T$ and $\mathbf R$ fit & 0 &
\padone{7.26} & \blankunr{\padone{4.96}} &
\blankunr{\padone{4.04}} &
\blankunr{\padone{1.02}} \\
\bottomrule
\end{tabular}
\end{table}

\begin{figure}[t]
\centering
\includegraphics[width=\columnwidth]{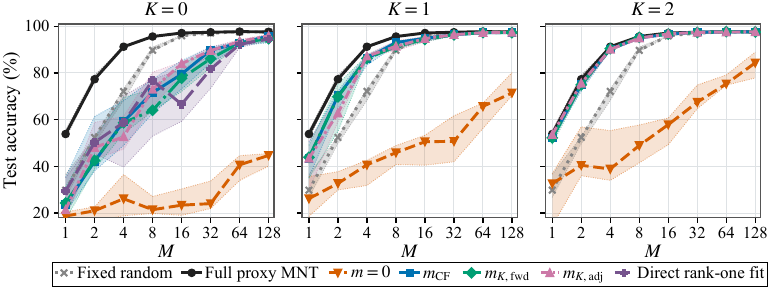}
\caption{Scene-classification test accuracy vs $M$ for $K\in\{0,1,2\}$. Curves and shading show medians and 10th--90th percentiles over five seeds, respectively.}
\label{fig:accuracy_vs_M}
\end{figure}

\textit{Scene-Classification Accuracy:}
We now ask whether the model-accuracy gains translate to classification-accuracy gains when DMA configurations and the classifier are optimized jointly.
The test accuracies in Fig.~\ref{fig:accuracy_vs_M} show that
$m_\mathrm{CF}$ with $K=2$ closely tracks training with the full proxy MNT across
$M$ (94.97\% vs. 95.57\% at $M=8$). At $K=0$, $m_\mathrm{CF}$ reaches
only 71.51\% with $M=8$, revealing a substantial task-level cost of
zeroth-order truncation. $m_{K,\mathrm{fwd}}^{\mathrm{(y)}}$ and $m_{K,\mathrm{adj}}^{\mathrm{(y)}}$ offer no consistent improvement over $m_\mathrm{CF}$, whereas the original parametrization  ($m=0$) performs
substantially worse. At $K=0$, the
directly fitted rank-one model does not consistently outperform $m_\mathrm{CF}$.

\section{Parametrization-Invariant MC-Strength Metric and MC-Unaware Benchmark Model}
\label{sec_MCmetric}

A common MC-unaware benchmark sets $\boldsymbol{\Gamma}=\mathbf 0$.
Yet M{\"o}bius reparametrizations preserve the end-to-end response
while changing $\boldsymbol{\Gamma}$ and thus the prediction obtained
with $\boldsymbol{\Gamma}=\mathbf 0$. The MC-unaware benchmark error therefore depends on the chosen MNT parametrization and can become arbitrarily large as
$\mathbf I_{N_{\mathrm M}}-m\boldsymbol\Gamma$ approaches a singularity.
Because setting $\boldsymbol{\Gamma}=\mathbf 0$ does not yield a
unique benchmark, we suggest defining the MC-unaware benchmark instead by directly
fitting the rank-one affine model in~\eqref{eq:direct-k0} to the
end-to-end response. Its minimum fitting error is independent of
the MNT parametrization and measures MC-induced nonlinearity,
although the fitted parameters need not be unique.
This parametrization-invariant metric quantifies the MC strength and was evaluated for binary controls in~\cite{rabault2024tacit}. For our strongly coupled
1-bit DMA, the joint $\mathbf T$-and-$\mathbf R$ fit yields
$\zeta_{\mathrm R}=6.86$\,dB and $\zeta_{\mathrm T}=5.89$\,dB,
as reported in Table~\ref{tab:raw_zeta}.
In the case of multibit controls, the model from~\eqref{eq:direct-k0} can be generalized by replacing $v_n\mathbf u_n\mathbf z_n^{\mathsf T}$ with
$g_n(v_n)\mathbf u_n\mathbf z_n^{\mathsf T}$, where $g_n$ is a scalar
function of the element's load state.

\section{Conclusion}

To summarize, we demonstrated that a simple closed-form M{\"o}bius reparametrization enables accurate low-order Neumann approximations of both forward and adjoint MNT evaluations for PMs, even under strong MC. 
For an experimentally calibrated proxy MNT model of a DMA with strong MC, our closed-form-reparametrized second-order approximation achieves forward and control-gradient accuracies of $\zeta_{\mathbf y}=21.68$\,dB and $\zeta_{\mathrm{VJP}}=21.70$\,dB, respectively, while reducing computational time by about one third and saved-tensor memory by a factor of ten. In a case study on end-to-end-optimized sensing, our reparametrized second-order-truncated MNT model closely approaches full-MNT performance ($95.0\%$ vs. $95.6\%$ at $M=8$). Moreover, the best achievable zeroth-order model accuracy provides a parametrization-invariant MC-unaware benchmark and MC-strength metric.

\bibliographystyle{IEEEtran}


\end{document}